\RequirePackage{fix-cm}
\documentclass[smallextended]{svjour3}       
\smartqed  
\usepackage{graphicx}
\usepackage{amsmath}
\usepackage{amsfonts}
\usepackage{amssymb}

\begin{document}

\title{Effective Field Theories in a Finite Volume
}


\author{A. Mart\'inez Torres\and S. Prelovsek\and E. Oset\and A. Ramos
}


\institute{A. Mart\'inez Torres\at
              Instituto de F\'isica, Universidade de S\~ao Paulo, Rua do Mat\~ao 1371, Butant\~a, CEP 05508-090, São Paulo, S\~ao Paulo,Brazil.
              \email{amartine@if.usp.br}           
                       \and
             S. Prelovsek
             \at 
             Instit\"ut f\"ur Theoretische Physik, Universit\"at Regensburg, D-93040 Regensburg, Germany. \\
\at Faculty of Mathematics and Physics, University of Ljubljana, 1000 Ljubljana, Slovenia. \\
\at Jozef Stefan,Institute, 1000 Ljubljana, Slovenia.\\
\and 
E. Oset
\at
Departamento de F\'isica Te\'orica and IFIC,Centro Mixto Universidad de Valencia-CSIC Institutos de Investigaci\'on de Paterna, Aptdo. 22085, 46071 Valencia, Spain.
\and 
A. Ramos
\at 
Departament d'Estructura i Constituents de la Mat\`eria and Institut de Ci\`encies del Cosmos, Universitat de Barcelona, Mart\'i i Franqu\`es 1, E-08028 Barcelona, Spain.
}

\date{Received: date / Accepted: date}

\maketitle

\begin{abstract}
In this talk I present the formalism we have used to analyze Lattice data on two meson systems by means of effective field theories. In particular I present the results obtained from a reanalysis of the lattice data on the $KD^{(*)}$ systems, where the states $D^*_{s0}(2317)$ and $D^*_{s1}(2460)$ are found as bound states of $KD$ and $KD^*$, respectively. We confirm the presence of such states in the lattice data and determine the contribution of the $KD$ channel in the wave function of $D^*_{s0}(2317)$ and that of $KD^*$ in the wave function of $D^*_{s1}(2460)$. Our findings indicate a large meson-meson component in the two cases.
\end{abstract}

\section{Introduction}
As it is well-known, Quantum Chromodynamics (QCD) is the fundamental theory which describes the strong interaction of hadrons and while at high energies the theory becomes perturbative and has been successfully tested by the experiment, the situation is very different in the low and intermediate energy regime. In this domain, QCD becomes non-perturbative and the extraction of information about the strong interaction of hadrons becomes more complicated and the use of non perturbative schemes becomes mandatory. 

Lattice QCD provides the only ab-initio framework to study the interaction of hadron systems at the low and intermediate energy region. In this case, the spacetime is discretized in a finite volume and energy eigenstates for the system under study are obtained by calculating two-point correlation functions considering interpolating operators resembling quark-antiquark, tetraquarks, meson-meson structures, etc., with the proper quantum numbers. Using the L\"uscher formalism, phase shifts at infinite volume can be obtained from the energy spectra found in the finite volume and information about the hadron system under consideration can be obtained, like scattering length, mass of the bound states/resonances formed, etc.~\cite{lu,dudek,morningstar,lang,bali}

Recently, an alternative method to the L\"uscher formalism has been developed. In this case the Lattice data on hadron systems is analyzed using effective field theories by means of an auxiliary kernel which constitute the input to the Bethe-Salpeter equation in a finite volume within a coupled channel formalism~\cite{doring,mar2,mar3,alba,guo}. Both method, up to exponentially suppressed terms, are equivalent~\cite{doring,guo}.

In this talk I present the results obtained for the $D^{(*)}K$, $D^{(*)}_s\eta$ systems by using effective field theories in a finite volume, in which the states $D^*_{s0}(2317)$ and $D^*_{s1}(2420)$ are formed as $D^{(*)}K$ bound states. Interestingly, there is lattice data available for the $D^{(*)}K$ systems~\cite{lang,bali} and here I present a reanalysis of the data of Ref.~\cite{lang}, where the L\"usher method was used to obtain information about the  $D^*_{s0}(2317)$ and $D^*_{s1}(2420)$ states.

\section{Formalism}
By discretizing the space in a cubic box of volume $V=L^3$, with $L$ being the length of one of the sides of the box, the Bethe-Salpeter equation can be written as~\cite{doring,mar2,mar3,alba,mar1}
\begin{equation}
\mathcal{T}(E,L)=[1-\mathcal{V}(E)\mathcal{G}(E,L)]^{-1}\mathcal{V}(E).\label{bse}
\end{equation}
In Eq.~(\ref{bse}), $E$ is the center of mass energy of the system, $\mathcal{V}(E)$ is the kernel, which is expressed as a matrix whose elements are the lowest order amplitudes describing the transitions between two meson-meson channels, let us call them $i$ and $j$, and $\mathcal{G}(E,L)$ is a matrix whose elements are the two meson loop function in the finite volume for the transition $i\to j$.

Particularly, we are going to analyze the lattice data of Ref.~\cite{lang} on the $D^{(*)}K$ system, thus, we need a parametrization of the kernel $\mathcal{V}(E)$ for these systems. A realistic parametrization can be found by following Refs.~\cite{daniel1,daniel2}, where using an effective field theory based on SU(4) symmetry, the kernel $\mathcal{V}(E)$ for the $D^{(*)} K$, $D^{(*)}_s\eta$ coupled channel system is given by 
\begin{align}
\mathcal{V}_{11}&=-\frac{1}{3f_\pi f_{D^{(*)}}}[\gamma (\bar t-\bar u)+s-\bar u+M^2_{D^{(*)}}+M^2_K],\nonumber\\
\mathcal{V}_{12}&=-\frac{1}{9f_\pi f_{D^{(*)}}}[\gamma(-s+2\bar t-\bar u)+2M^2_{D^{(*)}}+6M^2_K-4M^2_\pi],\label{V}\\
\mathcal{V}_{22}&=\frac{1}{6\sqrt{3}f_\pi f_{D^{(*)}}}[\gamma(\bar u-\bar t)-(3+\gamma)(s-\bar u)-M^2_{D^{(*)}}-3 M^2_K+2M^2_\pi],\nonumber
\end{align}
with $s=E^2$, the subscript 1 represents the $KD^{(*)}$ channel and the subscript 2 is associated with the $\eta D^{(*)}_s$ channel. The variables $\bar t$ and $\bar u$ correspond to the Mandelstam $t$ and $u$ variables, respectively, projected on $s$-wave, while $\gamma$ is a constant related to the SU(4) breaking of the model. The pion and $D^{(*)}$ decay constants are represented in Eq.~(\ref{V}) by $f_\pi$ and $f_{D^{(*)}}$, respectively. The expressions in Eq.~(\ref{V}), considering the dependence of $\bar t$ and $\bar u$ on the masses and the $s$ variable, suggest the following parametrization of the kernel $\mathcal{V}(E)$ for energies around the $D^{(*)}K$ threshold~\cite{mar1}
\begin{align}
\mathcal{V}_{ij}(E,\alpha,\beta)=\alpha_{ij}+\beta_{ij}(s-s_\text{th}), \quad s_\text{th}=(M_{D^{(*)}}+M_K)^2.\label{ker}
\end{align}
Although the exact values for $\alpha_{ij}$ and $\beta_{ij}$ are known for the theoretical model of Refs.~\cite{daniel1,daniel2}, when analyzing the lattice data we keep them as parameters to be determined by fitting the lattice data on the energy levels for the system. In this way we allow reasonable deviations from the kernels in Eq.~(\ref{V}) which are based on SU(4).

For the $\mathcal{G}(E,L)$ in Eq.~(\ref{bse}) we have 
\begin{align}
\mathcal{G}_{ij}(E,L)&=G_{ij}(E)+\lim_{q_\textrm{max}\to\infty}\left[\frac{1}{L^3}\sum_{q_r}^{q_\textrm{max}}I_i(\vec{q}_r)-\int\limits^{}_{q<q_\textrm{max}}\frac{d^3q}{(2\pi)^3}I_i(\vec{q}\,)\right]\delta_{ij},\nonumber\\
\vec{q}_r&= \frac{2\pi}{L}\vec{n}_r, ~~\vec{n}_r \in \mathbb{Z}^3\nonumber\\
G_{ij}(E)&=\int\frac{d^3q}{(2\pi)^3}I_{ij}(\vec{q}\,),\nonumber\\
I_i(\vec{q}\,)&=\frac{\omega_{1i}(\vec{q}\,)+\omega_{2i}(\vec{q}\,)}{2\omega_{1i}(\vec{q}\,)\omega_{2i}(\vec{q}\,)\left[s-(\omega_{1i}(\vec{q}\,)+
\omega_{2i}(\vec{q}\,))^2+i\epsilon\right]}\delta_{ij}\label{loop}.
\end{align}
In Eq.~(\ref{loop}), $G_{ij}$ is the two meson loop function in the infinite volume for the transition $i\to j$ and $\omega_{1i,2i}(\vec{q})=\sqrt{\vec{q}^2+m^2_{1i,2i}}$ is the on-shell energy of the mesons 1 and 2, respectively, constituting the channel $i$. This $G_i$ is divergent and needs to be regularized either with dimensional regularization or a cut-off. In the former case, subtraction constants are needed while in the latter a cut-off, $q^\prime_\text{max}$, is required.  In both cases, different values of the subtraction constants or $q^\prime_\text{max}$ produce changes in $G_i$ which can be reabsorbed in the parameters $\alpha_{ij}$ and $\beta_{ij}$ present in the kernel $\mathcal{V}$ (Eq.~(\ref{ker})). Thus, any reasonable value for the subtraction constants, typically around $-2$ for a regularization scale of 630 MeV, or $q^\prime_\text{max}$, typically of the order of 1000 MeV, can be used to regularize $G_i$ and the results obtained are basically independent of the subtraction constants or $q^\prime_\text{max}$.

Using the kernel $\mathcal{V}(E,\alpha,\beta)$ and loop function $\mathcal{G}(E,L)$ calculated in the finite volume $V=L^3$, the eigen-energies of the system under consideration can be obtained in the discretized space
from the resolution of
\begin{equation}
\textrm{det}[1-\mathcal{V}(E,\alpha,\beta)\mathcal{G}(E,L)]=0\label{det}
\end{equation}
for different values of $L$. The comparison between these levels and those obtained in a lattice study of the system determines the parameters $\alpha_{ij}$ and $\beta_{ij}$ through fitting the lattice data.

For the particular value of the parameters obtained from the fit, we can redefine $\mathcal{V}(E,\alpha,\beta)$ as $\mathcal{V}(E)$. Then, we can solve the Bethe-Salpeter equation at infinite volume using $\mathcal{V}(E)$ as kernel, obtaining in this way the scattering matrix $T$ at infinite volume as
\begin{align}
T(E)=[1-\mathcal{V}(E) G(E)]^{-1} \mathcal{V}(E).\label{bse2}
\end{align}
Bound states or resonances in the system considered appear as poles of the scattering matrix $T$ in the complex energy plane. From the $T$-matrix we can obtain information related to the nature of the resonance/bound state obtained by means of its residue: the residue of the scattering matrix determines the coupling $g_i$ of the states found to the different meson-meson channels considered when solving Eq.~(\ref{bse2}). These couplings satisfy the following sum rule~\cite{daniel3,hyodo}
\begin{align}
-\sum_i g^2_i\frac{dG_i}{ds}\Bigg|_{\text{pole}}=1-Z\label{g_i}.
\end{align}
Each of the terms inside the summation symbol of Eq.~(\ref{g_i}) represents the probability of finding in the wave function of the state the meson-meson channel $i$, while the probability of finding any other component different to the meson-meson channel $i$ is given by the $Z$ function. 

\section{Results}
First, we ignore the $\eta D^{(*)}_s$ channel and consider the $KD^{(*)}$ channel as the only coupled channel. In such a case, the kernel $\mathcal{V}$ contains just the transition $KD^{(*)}\to K D^{(*)}$ and two parameters, $\alpha_{11}$ and $\beta_{11}$, need to be determined by solving Eq.~(\ref{det}) while fitting to the lattice data of Ref.~\cite{lang}. The results for the energy levels obtained are shown in Fig.~\ref{res}.
\begin{figure} 
\begin{center}
\includegraphics[width=0.45\textwidth,clip]{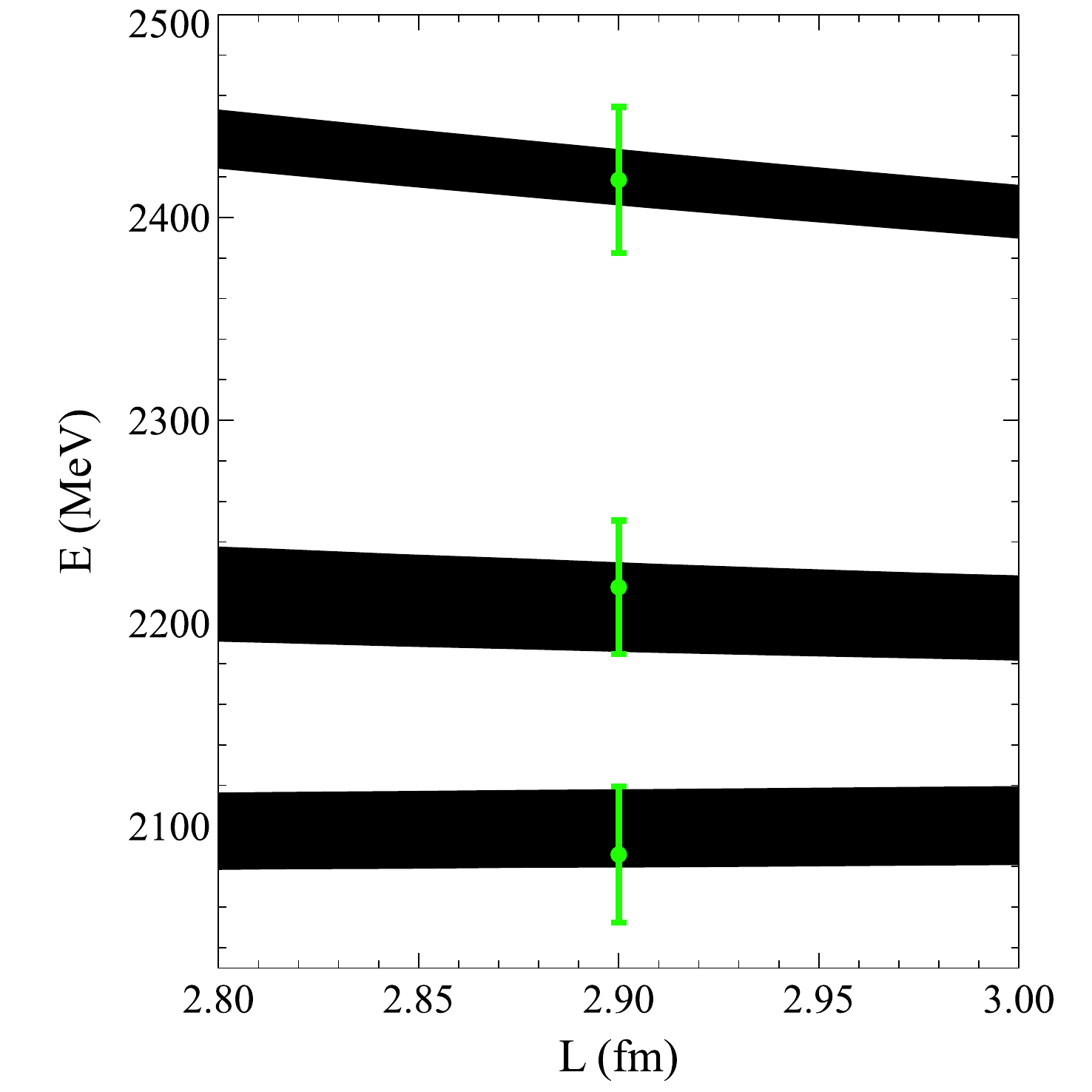}
\includegraphics[width=0.45\textwidth,clip]{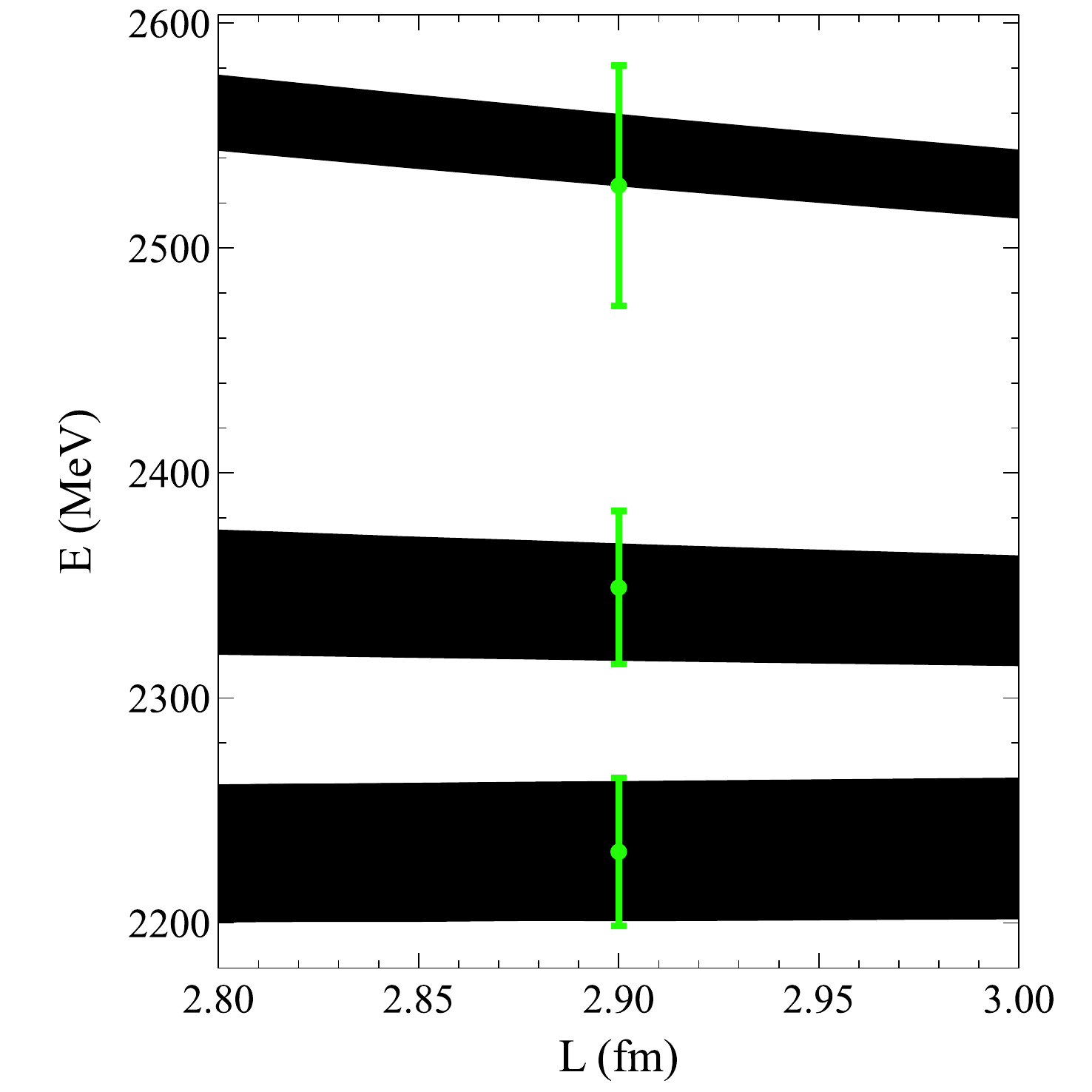}
\caption{Fits to the lattice data of Ref.~\cite{lang} for the $KD$ system (left panel) and the $KD^*$ system (right panel) which have been obtained when solving Eq.~(\ref{det}).}\label{res}
\end{center}
\end{figure}

Using the same kernel $\mathcal{V}$ to solve Eq.~(\ref{bse2}), we get the $KD^{(*)}$ scattering matrix $T$ in the infinite volume and using this $T$-matrix we can obtain the pole position for $D_{s0}(2317)$, $D^*_{s1}(2460)$ or binding energy
$B$ of the $KD^{(*)}$ systems, the probability $P$ of finding the $KD^{(*)}$ component in the respective wave function, the scattering length $a_0$ and the effective range $r_0$. The results are summarized in Table~\ref{res} and can be compared with the ones obtained in Ref.~\cite{lang} by means of the L\"usher method and the effective range formula, finding a good agreement. As can be seen in Table~\ref{res}, both $D_{s0}(2317)$ and $D^*_{s1}(2460)$ have a large $KD$ and $KD^*$ component, respectively, in their wave functions.

\begin{table}
\centering
\caption{Results obtained for $D_{s0}(2317)$ and $D^*_{s1}(2460)$ as bound states of the $KD$ and $KD^*$ systems, respectively.}\label{res}
\begin{tabular}{ccc}
&$KD$&$KD^*$\\
\hline
$B$ (MeV)&$46\pm 21$&$52\pm 22$\\
$P$ (\%)&$76\pm 12$&$53\pm 17$\\
$a_0$ (fm)&$1.2\pm 0.6$&$-0.9\pm 0.3$\\
$r_0$ (fm)&$0.04\pm 0.16$&$0.3\pm 0.4$\\
\end{tabular}
\end{table}

However, in the effective field theories at infinite volume, the states $D^*_{s0}(2317)$ and $D^*_{s1}(2460)$ are not pure $KD^{(*)}$ bound states, and there is a non negligible $\eta D^{(*)}_s$ component, representing around $20\%$ of the wave function of the state~\cite{daniel1,daniel2}. It would be then interesting to quantify from the lattice data of Ref.~\cite{lang} the relevance of the $\eta D_s$ component in the wave function of $D^*_{s0}(2317)$ and that of the $\eta D^*_s$ component in case of $D^*_{s1}(2460)$.  With this purpose, we can solve Eq.~(\ref{det}) considering now $KD^{(*)}$ and $\eta D^{(*)}_s$ as coupled channels. Since $\mathcal{V}_{21}=\mathcal{V}_{12}$, we have then 6 parameters to be determined by fitting the data of Ref.~\cite{lang}, but we have just three data points. Thus, the only way we can fit the data of Ref.~\cite{lang} with two coupled channels is by using energy independent kernels, i.e., 
\begin{align}
\mathcal{V}(\alpha)&=\left(\begin{array}{cc}\mathcal{V}_{11}(E,\alpha_{11},0)&\mathcal{V}_{12}(E,\alpha_{12},0)\\\mathcal{V}_{12}(E,\alpha_{12},0)&\mathcal{V}_{22}(E,\alpha_{22},0)\end{array}\right)=\left(\begin{array}{cc}\alpha_{11}&\alpha_{12}\\\alpha_{12}&\alpha_{22}\end{array}\right),\label{Valpha}\\
\mathcal{G}(E,L)&=\left(\begin{array}{cc}\mathcal{G}_{11}(E,L)&0\\0&G_{22}(E,L)\end{array}\right).
\end{align}
In this way we have three parameters to be determined, $\alpha_{11}$, $\alpha_{12}$ and $\alpha_{22}$. In such a situation the wave function of the states $D^*_{s0}(2317)$ and $D^*_{s1}(2460)$ would be saturated 
with the $KD^{(*)}$ and $\eta D^{(*)}_s$ channels. This is due to the relation between the energy dependence of the kernel and the $Z$ function present in  Eq.~(\ref{g_i}), as shown in Ref.~\cite{hyodo}. In this way, we can obtain the weight of $\eta D_s$ in the wave function of $D^*_{s0}(2317)$ and that of $\eta D^*_s$ in the wave function of $D^*_{s1}(2460)$ by comparing the probabilities found with just the $KD^{(*)}$ channel and an energy dependent kernel (see Table~\ref{res}) with those found with two coupled channels, $KD^{(*)}$ and $\eta D^{(*)}_s$, and constant kernel.

However, we do not find any suitable fit when trying to fit the data of Ref.~\cite{lang} by solving Eq.~(\ref{bse}) with the kernel in Eq.~(\ref{Valpha}). This result could be considered as an evidence that the energy levels obtained in Ref.~\cite{lang} do not have information on the $\eta D_s$ or $\eta D^{(*)}_s$ channels:  although in a dynamical lattice simulation all states with a given quantum number are in principle expected, a poor basis of interpolating fields is insufficient to render them in practice. In this sense, future lattice simulations of the $KD^{(*)}$ systems should consider explicitly $\eta D^{(*)}_s$ interpolators to shed more light on the nature of $D^*_{s0}(2317)$ and $D^*_{s1}(2460)$.

\section{Conclusions}
In this talk I have presented the results that we have found from a reanalysis of the lattice spectra obtained in Ref.~\cite{lang} for the $KD^{(*)}$ systems. Our analysis confirms the existence of $D^*_{s0}(2317)$ and $D^*_{s1} (2460)$ with a large $KD$ component in the wave function of $D^*_{s0}(2317)$ and $KD^*$ component in the wave function of $D^*_{s1} (2460)$. Our analysis suggests that the future lattice simulations should explicitly include $\eta D^{(*)}_s$ interpolators, allowing in this way the determination of the probability of finding such components in the respective wave function of the states.

\begin{acknowledgements}
A.M.T gratefully acknowledges the financial support received from FAPESP (under  the
grant number 2012/50984-4) and the support from CNPq (under the grant number 310759/2016-1). This work is partly supported by the Spanish Ministerio de Economia y Competitividad and European FEDER funds under contract numbers FIS2011-28853-C02-01 and FIS2011-28853-C02-02,  by the Generalitat Valenciana in the program Prometeo II, 2014/068, and by Grant 2014SGR-401from the Generalitat de Catalunya. We acknowledge the support of the European Community-Research Infrastructure Integrating Activity Study of Strongly Interacting Matter (acronym
HadronPhysics3, Grant Agreement n. 283286) under the Seventh Framework Programme of EU. 
\end{acknowledgements}


\begin{thebibliography}{}
\bibitem{lu}
  M.~L\"uscher,Commun.\ Math.\ Phys. {\bf 105} (1986) 153, {\it idem} Nucl. Phys. {\bf B 354} (1991) 531.
\bibitem{dudek}
J. J. Dudek, Phys. Rev. {\bf D84} (2011) 074023.
\bibitem{morningstar}
C. Morningstar, J. Bulava, J. Foley, K. J. Juge, D. Lenkner, M.  Peardon,   and  C. H.  Wong, Phys.  Rev. {\bf D83} (2011) 114505. 
\bibitem{lang}
C.B. Lang, L. Leskovec, D. Mohler, S. Prelovsek and R.M. Woloshyn, Phys. Rev. {\bf D 90} (2014) 034510.
  \bibitem{bali} 
  G.~S.~Bali, S.~Collins, A.~Cox and A.~Sch\"afer, Phys.\ Rev.\ {\bf D96} (2017) 07450. 
\bibitem{doring}
M. Doring, U. G. Mei\ss ner, E. Oset and A. Rusetsky, Eur. Phys. J. {\bf A 47} (2011) 139, {\it idem} \emph{Eur. Phys. J.} {\bf A 48}  (2012) 114.
\bibitem{mar2}
A. Martinez Torres, L.R. Dai, C. Koren, D. Jido and E. Oset, Phys. Rev.{\bf D 85} (2012) 014027.
\bibitem{mar3}
A. Martinez Torres, M. Bayar, D. Jido and E. Oset, Phys. Rev. C {\bf 86}  (2012) 055201.
\bibitem{alba} 
  M.~Albaladejo, J.~A.~Oller, E.~Oset, G.~Rios and L.~Roca, JHEP {\bf 1208} (2012) 071.
  \bibitem{guo} 
  Z.~H.~Guo, L.~Liu, U.~G.~Mei\ss ner, J.~A.~Oller and A.~Rusetsky,
  Phys.\ Rev.\ D {\bf 95}, no. 5, 054004 (2017).
  \bibitem{mar1} 
  A.~Martinez Torres, E.~Oset, S.~Prelovsek and A.~Ramos,\emph{JHEP} {\bf 1505}, 153 (2015)
  \bibitem{daniel1} 
  D.~Gamermann, E.~Oset, D.~Strottman and M.~J.~Vicente Vacas, \emph{Phys.\ Rev.}\ {\bf D 76} (2007) 074016.
  \bibitem{daniel2} 
  D.~Gamermann and E.~Oset, \emph{Eur.\ Phys.\ J.}\ {\bf A 33} (2007) 119.
\bibitem{daniel3} 
  D.~Gamermann, J.~Nieves, E.~Oset and E.~Ruiz Arriola, \emph{Phys.\ Rev.} {\bf D 81} (2010) 014029.
\bibitem{hyodo} 
  T.~Hyodo, D.~Jido and A.~Hosaka, \emph{Phys.\ Rev.} {\bf C 78} (2008) 025203.

\end{thebibliography}
\end{document}